\documentclass[prc,aps,twocolumn,showpacs,amssymb,superscriptaddress,float]{revtex4}

\usepackage{amsmath}
\usepackage{graphicx}
\usepackage{dcolumn}
\usepackage{float}

\begin{document}

\title{Shell-model study of the $N=82$ isotonic chain with a realistic
  effective hamiltonian}

\author{L. Coraggio}
\affiliation{Istituto Nazionale di Fisica Nucleare, \\
Complesso Universitario di Monte  S. Angelo, Via Cintia - I-80126 Napoli,
Italy}
\author{A. Covello}
\affiliation{Istituto Nazionale di Fisica Nucleare, \\
Complesso Universitario di Monte  S. Angelo, Via Cintia - I-80126 Napoli,
Italy}
\affiliation{Dipartimento di Scienze Fisiche, Universit\`a
di Napoli Federico II, \\
Complesso Universitario di Monte  S. Angelo, Via Cintia - I-80126 Napoli,
Italy}
\author{A. Gargano}
\affiliation{Istituto Nazionale di Fisica Nucleare, \\
Complesso Universitario di Monte  S. Angelo, Via Cintia - I-80126 Napoli,
Italy}
\author{N. Itaco}
\affiliation{Istituto Nazionale di Fisica Nucleare, \\
Complesso Universitario di Monte  S. Angelo, Via Cintia - I-80126 Napoli,
Italy}
\affiliation{Dipartimento di Scienze Fisiche, Universit\`a
di Napoli Federico II, \\
Complesso Universitario di Monte  S. Angelo, Via Cintia - I-80126 Napoli,
Italy}
\author{T. T. S. Kuo}
\affiliation{Department of Physics, SUNY, Stony Brook, New York 11794}

\date{\today}

\begin{abstract}
We have performed shell-model calculations for the even- and odd-mass
$N=82$ isotones, focusing attention on low-energy states. 
The single-particle energies and effective two-body interaction have
been both determined within the framework of the time-dependent
degenerate linked-diagram perturbation theory, starting from a
low-momentum interaction derived from the CD-Bonn nucleon-nucleon
potential. In this way, no phenomenological input enters our effective
Hamiltonian,  whose reliability is evidenced by the good agreement
between theory and experiment.
\end{abstract} 


\maketitle

\section{Introduction}

The $N=82$ region has long been the subject of shell-model studies
owing to the strong doubly magic character of $^{132}$Sn. 
In this context, several calculations have been performed employing
purely phenomenological two-body effective interactions
\cite{Sarkar01}. 
Clearly, a main issue in the study of this region is that it provides
the opportunity to investigate the effect of adding protons to a
doubly-magic core over a large number of nuclei (from mass number
A=133 to A=152). 
Comprehensive studies of this kind were conducted some twenty years
ago in Refs. \cite{Andreozzi90} and \cite{Wildenthal90}. 
In the former a simple pairing Hamiltonian and empirical
single-particle (SP) energies were used, while in the latter the
two-body matrix elements and single-particle energies were determined
simultaneously by a least squares fit to the ground-state and
excited-state energies drawn from all experimentally studied $N=82$
nuclei.
 
Starting in the mid 1990s, however, shell-model calculations employing
effective interactions derived from realistic nucleon-nucleon ($NN$)
potentials have been performed in the $N=82$ region, which have
generally yielded very good results
\cite{Andreozzi97,Covello97,Holt97,Suhonen98,Covello02}.
Some of these studies \cite{Covello97,Holt97,Suhonen98} have focused
attention on the evolution of the low-energy properties of the $N=82$
isotones as a function of $A$, showing that the observed behavior is
well reproduced by the theory.
In Refs. \cite{Holt97,Suhonen98} the shell-model results were also
compared with those of a quasiparticle random-phase approximation
calculation leading to the conclusion that the  low-lying states are
equally well described by these two approaches.

A main merit of the above mentioned realistic shell-model calculations
is, of course, that no adjustable parameter appears in the matrix
elements of the two-body effective interaction. 
In all of them, however, the single-particle energies have been taken
from experiment. 
Actually, the calculational techniques used in these studies are based
on the time-dependent degenerate linked-diagram perturbation theory
\cite{Kuo90}, which provides an effective shell-model hamiltonian,
$H_{\rm eff}$, containing both one- and two-body components.
Usually, however, the one-body components, which represent the
theoretical SP energies, are subtracted from $H_{\rm eff}$ and
replaced by those obtained from the experimental spectra of nuclei
with one valence nucleon \cite{Shurpin83}.
This procedure has indeed led to a very good description of nuclear
properties in different mass regions \cite{Coraggio09}.

Recently, we have performed fully realistic
shell-model calculations employing both theoretical SP energies and
two-body interactions in studies of $p-$ and $sd-$shell nuclei
\cite{Coraggio05c,Coraggio07b}.
In these studies we have renormalized the high-momentum repulsive
components of the bare $NN$ potential $V_{NN}$ by way of the so-called
$V_{\rm low-k}$ approach \cite{Bogner01,Bogner02}, which provides a
smooth potential preserving exactly the onshell properties of the
original $V_{NN}$ up to a cutoff $\Lambda$.
The effective hamiltonian has then been derived within the framework
of the time-dependent degenerate linked-diagram perturbation theory.
In both papers \cite{Coraggio05c,Coraggio07b} a cutoff momentum
$\Lambda=2.1$ fm$^{-1}$ was employed. 
This corresponds to the laboratory energy $E_{lab} \simeq 350$ MeV,
which is the inelastic threshold of the $NN$ scattering.
The results of these studies compare well with experiment and other
shell-model calculations \cite{Cohen65,Richter06}.

On these grounds, we have found it interesting to carry out a similar
shell-model study of the $N=82$ chain, focusing attention on the
low-lying states of both even- and odd-mass isotones.
In this case, the SP energies and two-body effective interaction are
determined from a $V_{\rm low-k}$ derived from the high-precision
CD-Bonn potential \cite{Machleidt01b} with a cutoff momentum
$\Lambda=2.6$ fm$^{-1}$.
This value of $\Lambda$, which is somewhat larger than that used in
our studies of the light nuclei, is needed to obtain a reasonable
description of the experimental SP spectrum of the medium-heavy
nucleus $^{133}$Sb.
This is mainly due to the fact that, at variance with the model spaces
for light nuclei, in this case the 50-82 major shell contains the
intruder $0h_{11/2}$ state, whose relative energy turns out to be
strongly sensitive to the value of $\Lambda$. 
More precisely, for small values of $\Lambda$ this state lies far away
from the other SP levels of the 50-82 shell, while the correct
structure of this shell is restored when increasing the value of
$\Lambda$ \cite{Coraggio08b}.

The dependence of the results on $\Lambda$ would obviously vanish when
complementing the $V_{\rm low-k}$ effective two-body interaction with
different three- and higher-body components for each value of $\Lambda$.
However, it is at present a very hard task, from the computational
point of view, to take into account two- and three-body forces on an
equal footing in a third-order perturbative calculation for medium-
and heavy-mass nuclei.
Therefore, the value of the cutoff $\Lambda$ may be considered as a
parameter that we have chosen so as to reproduce satisfactorily the
experimental SP spectrum.
For the sake of completeness, it should be mentioned that the trend of
our calculated SP energies suggests that a larger value of the cutoff
could even improve the agreement with experiment.
However, this would amount to include higher-momentum components, which
would rapidly deteriorate the convergence properties of the perturbative
expansion used to derive the effective shell-model hamiltonian.
In the following section we show that our choice of $\Lambda$
is a reasonable compromise to assure both good perturbative
properties and quality of the results.

The paper is organized as follows. 
In Sec. II we give a brief outline of our calculations, focusing
attention on the convergence properties of the effective hamiltonian.
Sec. III is devoted to the presentation and discussion of our
results, while some concluding remarks are given in Sec. IV.

\section{Outline of calculations}
Our goal is to derive an effective hamiltonian for shell-model
calculations in the proton $2s1d0g0h$ shell, which is the standard
model space to describe the spectroscopic properties of $N=82$
isotones.
Within the framework of the shell model, an auxiliary one-body
potential $U$ is introduced in order to break up the hamiltonian for a
system of $A$ nucleons as the sum of a one-body term $H_0$, which
describes the independent motion of the nucleons, and a residual
interaction $H_1$:

\begin{eqnarray}
H & = & \sum_{i=1}^{A} \frac{p_i^2}{2m} + \sum_{i<j=1}^{A} V_{ij}^{NN} = T +
V = \nonumber \\
~ & = & (T+U)+(V-U) = H_{0}+H_{1}~~.
\label{smham}
\end{eqnarray}

Once $H_0$ has been introduced, a reduced model space is defined in
terms of a finite subset of $H_0$'s eigenvectors. 
The diagonalization of the many-body hamiltonian (\ref{smham}) in an
infinite Hilbert space, that it is obviously unfeasible, is then
reduced to the solution of an eigenvalue problem for an effective 
hamiltonian $H_{\rm eff}$ in a finite space.

In this paper, we derive $H_{\rm eff}$ by way of the time-dependent
perturbation theory \cite{Kuo90}.
Namely, $H_{\rm eff}$ is expressed through the Kuo-Lee-Ratcliff (KLR)
folded-diagram expansion in terms of the vertex function
$\hat{Q}$-box, which is composed of irreducible valence-linked
diagrams \cite{Kuo71,Kuo81}.
We include in the $\hat{Q}$-box one- and two-body Goldstone diagrams
through third order in $H_1$.
The folded-diagram series is summed up to all orders using the
Lee-Suzuki iteration method \cite{Suzuki80}.

The hamiltonian $H_{\rm eff}$ contains one-body contributions, whose
collection is the so-called $\hat{S}$-box \cite{Shurpin83}. 
As mentioned in the Introduction, in realistic shell-model
calculations it is customary to use a subtraction procedure, 
so that only the two-body terms of $H_{\rm eff}$,
which make up the effective interaction $V_{\rm eff}$, are retained
while the SP energies are taken from experiment.
In this work, we have adopted a different approach employing SP
energies obtained from the $\hat{S}$-box calculation.
In this regard, it is worth pointing out that, owing to the presence
of the $-U$ term in $H_1$, $U$-insertion diagrams arise in the 
$\hat{Q}$-box.
In our calculation we use the harmonic oscillator (HO) potential,
$U=\frac{1}{2} m \omega ^2 r^2$, and take into account all
$U$-insertion diagrams up to third order.
The oscillator parameter is $\hbar \omega = 7.88$ MeV, according to
the expression \cite{Blomqvist68} $\hbar \omega= 45 A^{-1/3} -25  A^{-2/3}$
 for $A=132$.

Let us now outline the $V_{\rm low-k}$ approach \cite{Bogner01,Bogner02} 
to the renormalization of $V_{NN}$. 
The repulsive core contained in $V_{NN}$ is smoothed by integrating
out the high-momentum modes of $V_{NN}$ down to a certain cutoff $\Lambda$. 
This integration is carried out with the requirement that the deuteron
binding energy and phase shifts of $V_{NN}$ are preserved by $V_{\rm
  low-k}$, which is achieved  by the following $T$-matrix equivalence
approach. 
We start from the half-on-shell $T$ matrix for $V_{NN}$ 
\begin{eqnarray}
 T(k',k,k^2) =  V_{NN}(k',k) +
 ~~~~~~~~~~~~~~~~~~~~~~~~~~~~~~~~~~~~~~~~ \nonumber \\ 
\mathcal{P} \int _0 ^{\infty} q^2 dq V_{NN} (k',q)
\frac{1}{k^2-q^2} T(q,k,k^2 ) ~~, ~~~~~~~~~~~~~~~~~~~~~
\end{eqnarray}

\noindent
where $\mathcal{P}$ denotes the principal value and  $k,~k'$, and $q$
stand for the relative momenta. 
The effective low-momentum $T$ matrix is then defined by
\begin{eqnarray}
T_{\rm low-k } (p',p,p^2) =  V_{\rm low-k }(p',p) + 
~~~~~~~~~~~~~~~~~~~~~~~~~~~~~~~ \nonumber \\ 
\mathcal{P} \int _0 ^{\Lambda} q^2 dq  V_{\rm low-k }(p',q) 
\frac{1}{p^2-q^2} T_{\rm low-k} (q,p,p^2) ~~,~~~~~~~~~~~
\end{eqnarray}

\noindent
where the intermediate state momentum $q$ is integrated from 0 to the
momentum space cutoff $\Lambda$ and $(p',p) \leq \Lambda$. 
The above $T$ matrices are required to satisfy the condition 
\begin{equation}
T(p',p,p^2)= T_{\rm low-k }(p',p,p^2) \, ; ~~ (p',p) \leq \Lambda \,.
\end{equation}

The above equations define the effective low-momentum interaction 
$V_{\rm low-k}$, and it has been shown \cite{Bogner02} that their
solution is provided by the KLR folded-diagram expansion
\cite{Kuo71,Kuo90} mentioned before.
In addition to the preservation of the half-on-shell $T$ matrix, which
implies preservation of the phase shifts, this $V_{\rm low-k}$
preserves the deuteron binding energy, since eigenvalues are preserved
by the KLR effective interaction. 
For any value of $\Lambda$, the $V_{\rm low-k}$ can be calculated very
accurately using iteration methods. 
Our calculation of $V_{\rm low-k}$ is performed by employing the
method proposed in \cite{Andreozzi96}, which is based on the
Lee-Suzuki similarity transformation \cite{Suzuki80}. 

\begin{figure}[H]
\begin{center}
\includegraphics[scale=0.3,angle=90]{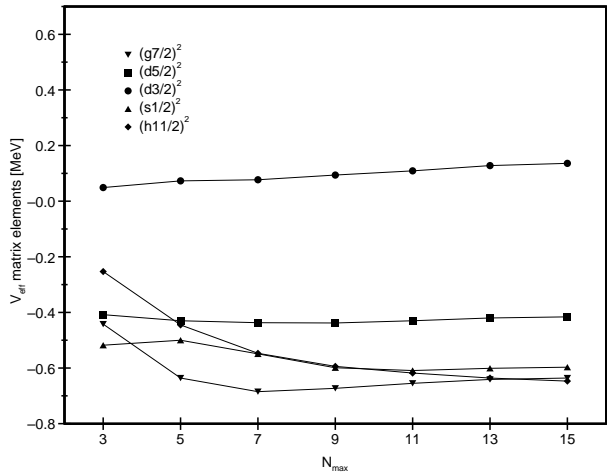}
\caption{Diagonal $J^{\pi}=0^+$ TBME of $V_{\rm eff}$
  as a function of $N_{\rm max}$ (see text for details).}
\label{Veffconv}
\end{center}
\end{figure}

As mentioned in the Introduction, our $V_{\rm low-k}$ has been derived
from the CD-Bonn $NN$ potential with a cutoff momentum $\Lambda=2.6$
fm$^{-1}$, and for protons the Coulomb force has been explicitly added
to $V_{\rm low-k}$.

A discussion of the convergence properties of $H_{\rm eff}$ is now in
order.
For the sake of clarity, we first consider the two-body matrix
elements (TBME) of $V_{\rm eff}$ and then the SP energies. 

\begin{table}[H]
\caption{$\hat{Q}$-box $J^{\pi}=0^+$ two-body matrix elements at third
order in $V_{\rm low-k}$ (in MeV) compared with those obtained by
calculating the Pad\'e approximant $[2|1]$.}
\begin{ruledtabular}
\begin{tabular}{ccc}
Configuration & 3rd order & Pad\'e $[2|1]$ \\
\colrule
$\langle (0g_{7/2})^2 | V_{Qbox} | (0g_{7/2})^2 \rangle$   & -0.8624 & -0.8757 \\
$\langle (0g_{7/2})^2 | V_{Qbox} | (1d_{5/2})^2 \rangle$   & -1.0022 & -1.0196 \\
$\langle (0g_{7/2})^2 | V_{Qbox} | (1d_{3/2})^2 \rangle$   & -0.6948 & -0.6999 \\
$\langle (0g_{7/2})^2 | V_{Qbox} | (2s_{1/2})^2 \rangle$   & -0.4642 & -0.4662 \\
$\langle (0g_{7/2})^2 | V_{Qbox} | (0h_{11/2})^2 \rangle$  &  2.0532 &  2.058  \\
$\langle (1d_{5/2})^2 | V_{Qbox} | (1d_{5/2})^2 \rangle$   & -0.7069 & -0.7138 \\
$\langle (1d_{5/2})^2 | V_{Qbox} | (1d_{3/2})^2 \rangle$   & -1.7611 & -1.8191 \\
$\langle (1d_{5/2})^2 | V_{Qbox} | (2s_{1/2})^2 \rangle$   & -0.6807 & -0.7555 \\
$\langle (1d_{5/2})^2 | V_{Qbox} | (0h_{11/2})^2 \rangle$  &  1.2028 &  1.2081 \\
$\langle (1d_{3/2})^2 | V_{Qbox} | (1d_{3/2})^2 \rangle$   &  0.0838 &  0.0447 \\
$\langle (1d_{3/2})^2 | V_{Qbox} | (2s_{1/2})^2 \rangle$   & -0.4817 & -0.4857 \\
$\langle (1d_{3/2})^2 | V_{Qbox} | (0h_{11/2})^2 \rangle$  &  1.0457 &  1.0459 \\
$\langle (2s_{1/2})^2 | V_{Qbox} | (2s_{1/2})^2 \rangle$   & -0.882  & -0.4686 \\
$\langle (2s_{1/2})^2 | V_{Qbox} | (0h_{11/2})^2 \rangle$  &  0.658  &  0.6586 \\
$\langle (0h_{11/2})^2 | V_{Qbox} | (0h_{11/2})^2 \rangle$ & -0.8025 & -0.8391 \\
\end{tabular}
\end{ruledtabular}
\label{table2b}
\end{table}

In Fig. \ref{Veffconv}, we report the diagonal $J^{\pi}=0^+$ TBME of
$V_{\rm eff}$ as a function of the
maximum allowed excitation energy of the intermediate states expressed
in terms of the oscillator quanta $N_{\rm max}$.
We have chosen the $J^{\pi}=0^+$ matrix elements because they are the
largest ones in the entire set of $V_{\rm eff}$ TBME and play a
key role in determining the relative spectra and ground-state
(g.s.) properties of the even $N=82$ isotones. 
From Fig. \ref{Veffconv} it is clear that our results have practically
achieved convergence at $N_{\rm max}=15$.

As regards the order-by-order convergence, an estimate  of the value
to which the perturbative series should converge may be obtained by
using Pad\'e approximants.
In Table \ref{table2b}, we compare all the $\hat{Q}$-box $J^{\pi}=0^+$
TBME calculated at third order with those given by the $[2|1]$
Pad\'e approximant \cite{Baker70}

\begin{equation}
[2|1] = V_{Qbox}^0 + V_{Qbox}^1 +
\frac{V_{Qbox}^2}{1-V_{Qbox}^{3}/V_{Qbox}^2}~~,
\end{equation}

\noindent
$V_{Qbox}^n$ being the $n$th-order contribution to the $J^{\pi}=0^+$
TBME in the linked-diagram expansion.
From Table \ref{table2b} we see that our third-order results are in good
agreement with those from the $[2|1]$ Pad\'e approximant, the
differences being all in the order of few tens of keV with the
exception of the diagonal matrix element for the $(2s_{1/2})^2$
configuration.
This indicates a weak dependence of our results on 
higher-order $\hat{Q}$-box perturbative terms.

\begin{figure}[H]
\begin{center}
\includegraphics[scale=0.4,angle=0]{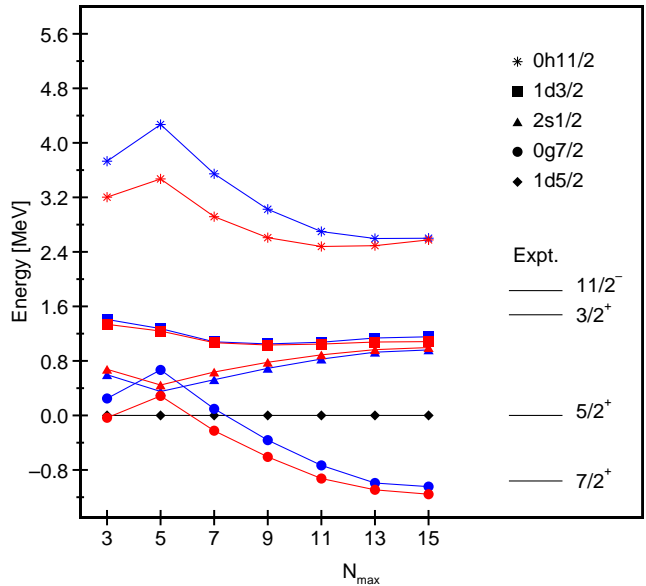}
\caption{(Color online) Theoretical relative SP energies as a function
  of $N_{\rm max}$ (see text for details). The experimental spectrum of
  $^{133}$Sb is also reported.}
\label{Sboxconv}
\end{center}
\end{figure}

In Fig. \ref{Sboxconv} we report the relative SP energies calculated
at third order in $H_1$ as a function of $N_{\rm max}$.
In the same figure, the results obtained by calculating the Pad\'e
approximant $[2|1]$ are also reported (red lines).
Once again, the agreement between third-order results and those from
this approximant points to a good perturbative behavior of the
calculated relative SP energies.
We also see that an $N_{\rm max}=15$ calculation yields satisfactorily
converged values for the relative SP energies.

A different situation occurs for the absolute SP
energies.
Actually, the calculated energy of the $1d_{5/2}$ level relative to
$^{132}$Sn is -4.468, -5.225, and -5.954 MeV for $N_{\rm max}$=11, 13,
and 15, respectively.
This shows that there is no sign of convergence of the absolute
SP energies as a function of the number of intermediate states.
Moreover, the Pad\'e approximant $[2|1]$ of the $1d_{5/2}$ energy for
$N_{\rm max}=15$ is equal to -7.262, showing that also the
order-by-order convergence of the absolute SP energies is
unsatisfactory.
This poor perturbative behavior may be traced to the chosen value of
the  cutoff momentum, $\Lambda=2.6$ fm$^{-1}$, which is substantially
larger than the standard one, $\Lambda \simeq 2.1$ fm$^{-1}$.
However, the inaccuracy of the absolute SP energies affects only the
g.s. energies of the nuclei considered.
In fact, the calculated spectroscopic properties (energy spectra,
electromagnetic transition rates) are certainly reliable, based on the
good convergence of the relative SP energies and TBME.

\section{Results and discussion}
We have performed calculations, using the Oslo shell-model
code \cite{EngelandSMC}, for the even-mass $N=82$ isotones up to
$^{154}$Hf, which is the last known nucleus belonging to this chain,
and for the odd-mass ones up to $^{149}$Ho, the last odd-mass nucleus
with a well-established experimental low-lying energy spectrum.
This is the longest isotonic chain approaching the proton drip line,
and therefore provides an interesting laboratory to study the
evolution of nuclear structure when adding pairs of identical
particles.

\begin{table}[H]
\caption{Calculated SP relative energies (in MeV), compared with the
  experimental spectrum of $^{133}$Sb \cite{nndc}. The values in
  parenthesis are the absolute SP energies with respect to the doubly
  closed $^{132}$Sn.}
\begin{ruledtabular}
\begin{tabular}{ccc}
$nlj$ & Calc. & Expt. \\
\colrule
$0g_{7/2}$  & 0.000 (-6.999) & 0.000 (-9.663) \\
$1d_{5/2}$  & 1.045 & 0.962 \\
$1d_{3/2}$  & 2.200 & 2.440 \\
$2s_{1/2}$  & 2.006 & ~~ \\
$0h_{11/2}$ & 3.645 & 2.793 \\
\end{tabular}
\end{ruledtabular}
\label{spetab}
\end{table}

In Table \ref{spetab} our calculated SP energies are reported and
compared with the experimental spectrum of $^{133}$Sb.
We see that the latter is on the whole reasonably well reproduced by
the theory.
However, while the calculated positions of the $1d_{5/2}$ and
$1d_{3/2}$ levels come very close to the experimental ones, the energy
of the $0h_{11/2}$ level is overestimated by about 0.85 MeV.
As we shall discuss later in this section, the energy of the
$2s_{1/2}$ level, for which there is no experimental information,
appears to be somewhat underestimated.

\begin{figure}[H]
\begin{center}
\includegraphics[scale=0.6,angle=0]{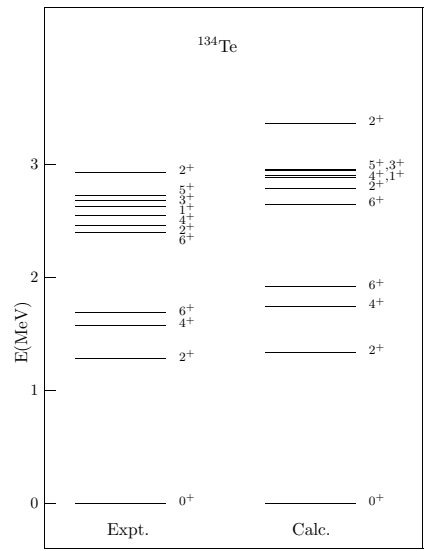}
\caption{Experimental and calculated $^{134}$Te spectra.}
\label{134te}
\end{center}
\end{figure}

A strong test for our effective hamiltonian is given by the
calculation of the energy spectrum of $^{134}$Te, since the theory of
the effective interaction is tailored for systems with two valence
nucleons.
From Fig. \ref{134te}, where the experimental \cite{nndc} and
calculated $^{134}$Te spectra are reported up to 3.5 MeV excitation
energy, we see that a very good agreement is indeed obtained.

\begin{figure}[H]
\begin{center}
\includegraphics[scale=0.4,angle=0]{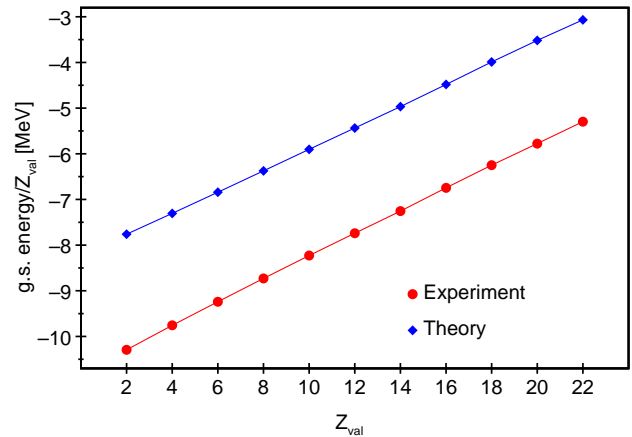}
\caption{(Color online)  Experimental and calculated ground-state 
energies per valence proton for $N=82$ isotones from $A=134$ to
154. $Z_{\rm val}$ is the number of valence protons.}
\label{gndsten}
\end{center}
\end{figure}

In Fig. \ref{gndsten}, we show the calculated and experimental
\cite{Audi03} g.s. energies (relative to the $^{132}$Sn core) per
valence proton as a function of the number of valence particles
$Z_{\rm val}$ of even-mass isotopes. 
We see that the experimental and theoretical curves are practically
straight lines having the same slope, while being about 2.4 MeV
apart.
This discrepancy is essentially the same as that existing between  the
theoretical and experimental g.s. energies of $^{133}$Sb (see Table
\ref{spetab}).
This confirms the reliability of our SP spacings and TBME, since the
pattern of the theoretical curve depends only on these quantities.

\begin{figure}[H]
\begin{center}
\includegraphics[scale=0.3,angle=90]{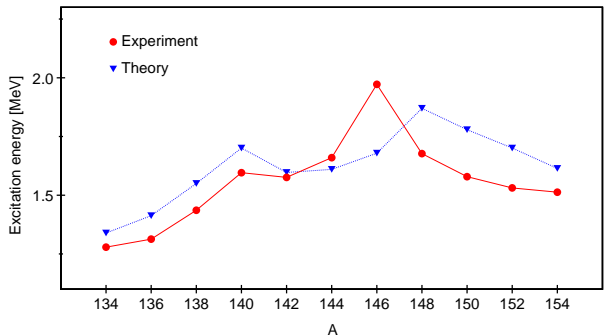}
\caption{(Color online) Experimental and calculated excitation energies 
of the yrast $J^{\pi}=2^+$ states for $N=82$ isotones.}
\label{J2p}
\end{center}
\end{figure}

\begin{figure}[H]
\begin{center}
\includegraphics[scale=0.3,angle=90]{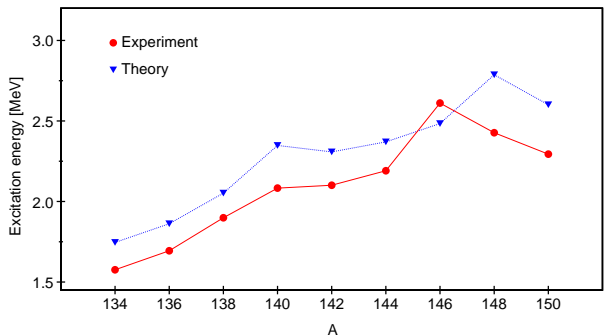}
\caption{(Color online) Experimental and calculated excitation energies 
of the yrast $J^{\pi}=4^+$ states for $N=82$ isotones.}
\label{J4p}
\end{center}
\end{figure}

\begin{figure}[H]
\begin{center}
\includegraphics[scale=0.3,angle=90]{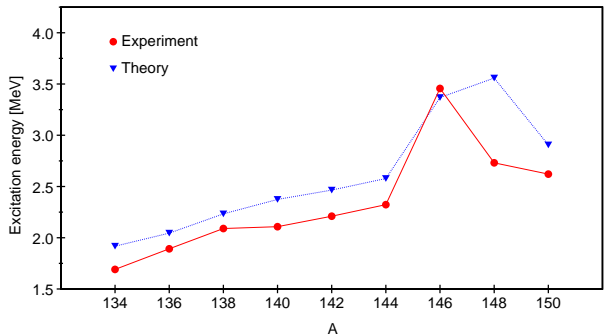}
\caption{(Color online) Experimental and calculated excitation energies 
of the yrast $J^{\pi}=6^+$ states for $N=82$ isotones.}
\label{J6p}
\end{center}
\end{figure}

In Figs. \ref{J2p}, \ref{J4p}, and \ref{J6p} we report, as a function
of the mass number $A$, the experimental and calculated excitation
energies of the $2^+$, $4^+$, and $6^+$ yrast states, respectively.
The experimental behavior is well reproduced, expecially for the
$J^{\pi}=2^+$ and $4^+$ states, for which the discrepancies do not
exceed 350 keV.
However, it should be noted that according to our calculations the
proton subshell closure, which experimentally occurs at $^{146}$Gd, is
not reproduced.
As mentioned before, this can be traced to the theoretical position of
the $2s_{1/2}$ orbital, which appears to be too low in energy. 
This is confirmed by the fact \cite{Andreozzi97} that this level has
to be placed at 2.8 MeV in order to reproduce the experimental energy
of the $J^{\pi}=\frac{1}{2}^+$ at 2.150 MeV in $^{137}$Cs, which is
predominantly of SP nature \cite{Wildenthal71}.
We have verified that, if the $2s_{1/2}$ SP level is placed at
2.8 MeV, we obtain that  the $2^+_1$ excitation energy in $^{146}$Gd
raises from 1.716 MeV to 1.976 MeV, while in $^{148}$Dy it decreases
from 1.827 to 1.758 MeV, thus reproducing the observed subshell closure.

\begin{table}[H]
\caption{Experimental and calculated $B(E2;2^+_1 \rightarrow 0^+_1)$.
The reduced transition probabilities are expressed in W.u..}
\begin{ruledtabular}
\begin{tabular}{ccc}
Nucleus &  Calc. & Expt. \\
\colrule
$^{134}$Te & 5.5   & $6.3 \pm 2.0$ \\
$^{136}$Xe & 9.06  & $16.6 \pm 2.4$ \\
$^{138}$Ba & 11.1  & $10.8 \pm 0.5$ \\
$^{140}$Ce & 14.5  & $13.8 \pm 0.3$ \\
$^{142}$Nd & 16.26 & $12.03 \pm 0.22$ \\
$^{144}$Sm & 16.6 & $11.9 \pm 0.4$ \\
\end{tabular}
\end{ruledtabular}
\label{E2}
\end{table}

To have a more complete test of the theory, we have also calculated 
the $B(E2;2^+_1 \rightarrow 0^+_1)$ transition rates up to $^{144}$Sm 
employing an effective operator obtained at third order in
perturbation theory, consistently with the derivation of $H_{\rm eff}$.
Our results are reported and compared with the experimental data in
Table \ref{E2}. 
We  see that the agreement is quite good, providing evidence for the
reliability of our calculated effective operator which takes into
account microscopically core-polarization effects.

\begin{figure}[H]
\begin{center}
\includegraphics[scale=0.3,angle=90]{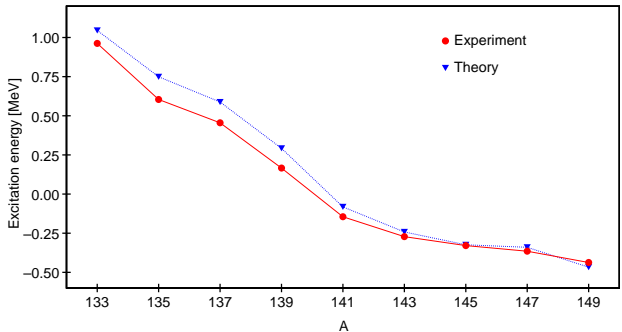}
\caption{(Color online) Experimental and calculated excitation
  energies of the yrast $J^{\pi}=\frac{5}{2}^+$ states relative to
  $J^{\pi}=(\frac{7}{2}^+)_1$ for odd-mass $N=82$ isotones.}
\label{J5p_odd}
\end{center}
\end{figure}

\begin{figure}[H]
\begin{center}
\includegraphics[scale=0.3,angle=90]{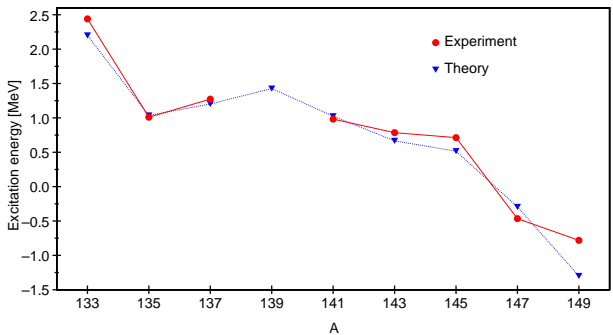}
\caption{(Color online) Experimental and calculated excitation
  energies of the yrast $J^{\pi}=\frac{3}{2}^+$ states relative to
  $J^{\pi}=(\frac{7}{2}^+)_1$ for odd-mass $N=82$ isotones.}
\label{J3p_odd}
\end{center}
\end{figure}

\begin{figure}[H]
\begin{center}
\includegraphics[scale=0.3,angle=90]{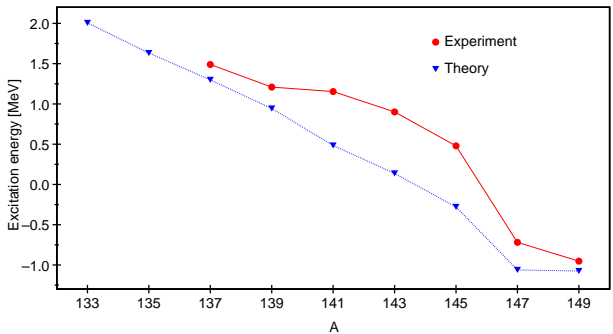}
\caption{(Color online) Experimental and calculated excitation
  energies of the yrast $J^{\pi}=\frac{1}{2}^+$ states relative to
  $J^{\pi}=(\frac{7}{2}^+)_1$ for odd-mass $N=82$ isotones.}
\label{J1p_odd}
\end{center}
\end{figure}

\begin{figure}[H]
\begin{center}
\includegraphics[scale=0.3,angle=90]{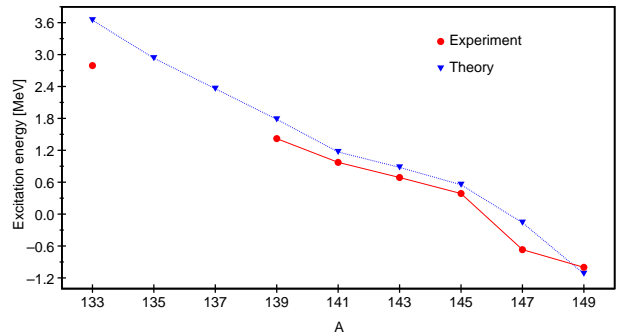}
\caption{(Color online) Experimental and calculated excitation
  energies of the yrast $J^{\pi}=\frac{11}{2}^-$ states relative to
  $J^{\pi}=(\frac{7}{2}^+)_1$ for odd-mass $N=82$ isotones.}
\label{J11m_odd}
\end{center}
\end{figure}

Let us now come to the odd-mass isotones. 
In Figs. \ref{J5p_odd}, \ref{J3p_odd}, \ref{J1p_odd}, and
\ref{J11m_odd} we report, as a function
of the mass number $A$, the experimental and calculated energies of
the $\frac{5}{2}^+$, $\frac{3}{2}^+$, $\frac{1}{2}^+$, and
$\frac{11}{2}^-$ yrast states relative to the $\frac{7}{2}^+$
yrast state, respectively.
The experimental behavior is well reproduced for the
$\frac{5}{2}^+,~\frac{3}{2}^+$, and $\frac{11}{2}^-$ states.
For the latter, however, most of the calculated energies are higher
than the experimental ones by about 300-400 keV, which reflects the
theoretical overestimation of the value of the $0h_{11/2}$ SP energy. 
Less satisfactory is the comparison between the calculated and
experimental behavior of the $\frac{1}{2}^+$ yrast states, which is a
further confirmation that the theoretical $2s_{1/2}$ level in
$^{133}$Sb lies too low in energy.
As a matter of fact, by performing a calculation within the seniority
scheme up to $v=3$, we have found that in $^{137}$Cs and $^{139}$La
the lowest $\frac{1}{2}^+$ states, which are quite well reproduced by
the theory, have significant seniority $v=3$ components.
From $A=141$ on, the experimental $(\frac{1}{2}^+)_1$ levels are
predominantly of SP nature, as testified by the spectroscopic factors
\cite{nndc}, and the corresponding calculated states, which are
dominated by $v=1$ components, underestimate their energies.

As pointed out in the Introduction, the main purpose of this study has
been to test the use of a fully realistic shell-model hamiltonian for
the description of medium-heavy mass nuclei. 
In this context, it should be noted that, aside from adopting
experimental single-particle energies, earlier calculations on the
N=82 isotones  differ from the present ones in other  respects, like
the starting $NN$ potential and the renormalization procedure. 
For instance, while we employ the CD-Bonn potential renormalized
through the  $V_{\rm low-k}$ procedure, in the works of
Refs. \cite{Holt97,Suhonen98} use is made of the Bonn A potential and
the Brueckner $G$-matrix approach.
A comparison between the results of earlier works and the present one
would not therefore be very meaningful. 
It is worth mentioning, however, that the agreement between experiment
and theory achieved in this paper is on the whole comparable with that
obtained in the previous calculations employing realistic effective
interactions.

\section{Concluding remarks}
As discussed in detail in the Introduction, the N=82 isotonic chain
has long been considered a benchmark for shell-model calculations.  
This has resulted in a number of theoretical works which have
practically all led to a good description of low-energy properties of
these nuclei, evidencing the strong doubly magic character of
$^{132}$Sn.

A main step towards a microscopic shell-model description of nuclear
structure has been the use of two-body effective interactions derived
from the free nucleon-nucleon potential. 
This approach, in which no adjustable parameter is involved in the
calculation of the TBME, has been successfully applied to the N=82
isotones in the last decade, showing the ability  of realistic
effective interactions to provide an accurate description of nuclear
structure properties. 

However, some important theoretical questions still remain open. 
In particular, these concern the calculation of the single-particle
energies and the role of three-body correlations. 
In the present paper, we have tried to investigate the former issue by
constructing a realistic effective shell-model hamiltonian for the
$N=82$ isotones, where both the SP energies and TBME have been
obtained starting from a $V_{\rm low-k}$ derived from the CD-Bonn
potential. 
This is a natural extension of our studies  of $p-$shell nuclei
\cite{Coraggio05c} and oxygen isotopes \cite{Coraggio07b}.
Here, we have shown that our effective hamiltonian yields results
which are in quite good agreement with experiment along the whole
chain of the $N=82$ isotones.
As regards the calculated single-particle energies, significant
discrepancies with the experimental values occur only for the
$0h_{11/2}$ and $2s_{1/2}$ levels. Note that for the latter no direct
comparison is possible since  the single-particle $2s_{1/2}$ state is
still missing in the experimental spectrum of $^{133}$Sb. 
However, we have verified that increasing the calculated value by 0.8
MeV the proton subshell closure occurs at $^{146}$Gd, in agreement
with the experimental findings.

At this point, the question naturally arises as to how contributions
from three-body forces would modify the present results, in particular
as regards the single-particle energies.
While this remains a major task for future investigations, we feel that
the present study paves the way toward a better understanding of the
microscopic foundations of the nuclear shell model.

\begin{acknowledgments}
This work was supported in part by the U.S. DOE Grant
No. DE-FG02-88ER40388.
\end{acknowledgments}

\bibliographystyle{apsrev}
\bibliography{biblio}
 
\end{document}